\documentclass[aps,prd,twocolumn,preprintnumbers, groupedaddress,nofootinbib,amssymb,notitlepage,eqsecnum]{revtex4-2}
\usepackage{bm}
\usepackage{amsmath,amsthm,amssymb}
\usepackage{amsfonts}
\usepackage{hyperref}
\usepackage{graphicx}

\allowdisplaybreaks[1]

\newcommand{\rd}{{\rm d}}
\newcommand{\be}{\begin{equation}}
\newcommand{\ee}{\end{equation}}
\newcommand{\ba}{\begin{eqnarray}}
\newcommand{\ea}{\end{eqnarray}}
\newcommand{\Mpl}{M_{\rm Pl}}

\begin{document}

\preprint{YITP-24-124, WUCG-24-09}

\title{Instability of nonsingular black holes in nonlinear electrodynamics}

\author{Antonio De Felice$^{a}$ and Shinji Tsujikawa$^{b}$}

\affiliation{$^{a}$Center for Gravitational Physics and Quantum Information, Yukawa Institute for Theoretical Physics, Kyoto University, 606-8502, Kyoto, Japan\\
$^{b}$Department of Physics, Waseda University, 3-4-1 Okubo, 
Shinjuku, Tokyo 169-8555, Japan}
 
\date{\today}

\begin{abstract}

We show that nonsingular black holes realized in nonlinear electrodynamics are always prone to Laplacian instability around the center because of a negative squared sound speed 
in the angular direction.
This is the case for both electric and magnetic BHs, where the instability of one of the vector-field perturbations leads to enhancing a dynamical gravitational perturbation in the even-parity sector. Thus, the background regular metric is no longer 
maintained in a steady state.
Our results suggest that the construction of stable, 
nonsingular black holes with regular 
centers, if they exist, requires 
theories beyond nonlinear electrodynamics.

\end{abstract}


\maketitle

\section{Introduction}
\label{Intro}

The vacuum black hole (BH) solutions predicted in General Relativity (GR) possess curvature singularities at their centers ($r=0$). 
Under several physical assumptions of spacetime and matter, 
Penrose showed that such singularities arise as an endpoint of the gravitational collapse \cite{Penrose:1964wq}.
However, the existence of singularity-free BHs is not 
precluded by relaxing some of these assumptions. 
For example, the nonsingular BH proposed by Bardeen \cite{Bardeen:1968} has a regular center due to the absence of global hyperbolicity of spacetime postulated in Penrose's theorem. Since quantum corrections to GR may manifest themselves in extreme gravity regimes, it is important to investigate whether curvature singularities can be eliminated in extended theories of gravity or matter.

It is known that nonlinear electrodynamics (NED) in the framework of GR allows the existence of spherically symmetric and static (SSS) BHs with regular centers \cite{Ayon-Beato:1998hmi, Ayon-Beato:1999kuh, Ayon-Beato:2000mjt, Bronnikov:2000vy, Dymnikova:2004zc, Ansoldi:2008jw, Balart:2014cga, Balart:2014jia, Fan:2016hvf, Rodrigues:2018bdc}. 
The Lagrangian ${\cal L}$ of NED depends on $F=-(1/4)F_{\mu \nu}F^{\mu \nu}$ nonlinearly, where $F_{\mu\nu}=\partial_{\mu}A_{\nu}-\partial_{\nu}A_{\mu}$ is the field strength of a covector field $A_{\mu}$. 
The action of Einstein-NED theory is given by 
\begin{equation}
{\cal S}=\int {\rm d}^{4}x \sqrt{-g} 
\left[ \frac{\Mpl^2}{2}R+{\cal L}(F) \right] \,,
\label{action}
\end{equation}
where $g$ is the determinant of a metric tensor $g_{\mu\nu}$, $\Mpl$ is the reduced Planck mass, and $R$ is the Ricci scalar. 
For example, Euler-Heisenberg theory \cite{Heisenberg:1936nmg} in quantum electrodynamics has a low-energy effective Lagrangian ${\cal L}=F+\alpha F^2$, where $\alpha F^2$ is a correction to the Maxwell term $F$. 
NED also accommodates Born-Infeld theory \cite{Born:1934gh} with the Lagrangian ${\cal L}=\mu^4 [1-\sqrt{1-2F/\mu^4}]$, 
in which the electron's self-energy is nondivergent by the finiteness of $F$. 
These subclasses of NED, when coupled with GR, give rise to hairy BH solutions \cite{Yajima:2000kw, Fernando:2003tz, Cai:2004eh, Dey:2004yt}, but there are in general curvature singularities at $r=0$ unless the functional form of ${\cal L}(F)$ is further extended.

The common procedure for realizing nonsingular BHs in NED is to assume the existence of regular metrics and reconstruct the Lagrangian ${\cal L}$ from the field equations 
of motion \cite{Ayon-Beato:1998hmi}.
In particular, for the magnetic BH, one can directly express ${\cal L}$ as a function of $F$ \cite{Ayon-Beato:2000mjt}. 
Nonsingular electric BHs constructed in this manner have finite values of $F$ and the electric field everywhere. 
For magnetic BHs, there is the divergence of $F$ at $r=0$, but the force exerted on a charged test particle is finite at any distance $r$ including $r=0$ \cite{Bronnikov:2000vy}. 
Nonsingular BHs can be designed to meet standard energy conditions, but not all regular solutions do \cite{Maeda:2021jdc}. For instance, violation of dominant energy conditions occurs for some regular BHs \cite{Bardeen:1968, Hayward:2005gi}.

Thus, at the background level, there are consistent regular BHs in NED evading Penrose's theorem. To see whether these BHs do not suffer from theoretical pathologies, we need to address their linear stability by analyzing perturbations on the SSS background. The BH perturbations in NED were studied in Refs.~\cite{Moreno:2002gg, Toshmatov:2018tyo, Toshmatov:2018ell, Toshmatov:2019gxg, Nomura:2020tpc} by focusing on the stability outside the outer horizon. 
It was found that there are viable parameter spaces in which the nonsingular BHs are plagued by neither ghosts nor Laplacian instabilities. However, the BH stability inside the horizon, especially around its center, is still unclear and has not been investigated to our best knowledge. 
In this letter, we will show that all nonsingular BHs in NED, including both electric and magnetic ones, are unstable due to the angular Laplacian instability around $r=0$.

\section{Nonsingular BHs in NED}
\label{backsec}

The line element on the SSS background is given by 
\begin{equation}
\rd s^{2} =-f(r) \rd t^{2} +h^{-1}(r)\rd r^{2} + 
r^{2} (\rd \theta^{2}+\sin^{2}\theta\, \rd \varphi^{2})\,,
\label{metric}
\end{equation}
where $f$ and $h$ are functions of the radial distance $r$. 
We consider the following covector-field configuration 
\be
A_{\mu}=\left[ A_0(r), 0, 0, -q_M \cos \theta \right]\,,
\label{Amu}
\ee
where $A_0$ is a function of $r$, and $q_M$ is 
a constant corresponding to a magnetic charge. 
Since the theory (\ref{action}) has $U(1)$ gauge 
invariance, $A_\mu\to A_\mu+\partial_\mu\chi_{\rm gauge}$, 
the longitudinal component $A_1(r)$ has been eliminated 
by choosing the gauge field as $\chi_{\rm gauge}(r)=-\int^r A_1(\rho) {\rm d}\rho$.

Varying the action (\ref{action}) with respect 
to $A_0$, $f$, and $h$, it follows that 
\ba
\left( \sqrt{h/f}\,r^2 {\cal L}_{,F} A_0' 
\right)' &=& 0\,,
\label{back1}\\
h'-\frac{1-h}{r} &=&
\frac{r}{\Mpl^2 f}
\left( f {\cal L}-h A_0'^2 {\cal L}_{,F} \right),
\label{back2}\\
\frac{f'}{f}-\frac{h'}{h} &=& 0\,,
\label{back3}
\ea
where a prime represents the derivative 
with respect to $r$, and 
${\cal L}_{,F} \equiv {\rm d}{\cal L}/{\rm d}F$.
The explicit form of $F$ is given by 
\be
F=\frac{h A_0'^2}{2f}-\frac{q_M^2}{2r^4}\,.
\label{Fexp}
\ee
From Eq.~(\ref{back3}), we obtain $f=Ch$, where $C$ is a constant. 
Using time reparametrization invariance, we can impose $f \to 1$ as $r \to \infty$, whereas the asymptotic flatness sets 
$h \to 1$ at spatial infinity.
Then we have $C=1$, so that  
\be
f=h\,.
\ee
With this condition, 
Eqs.~(\ref{back1}) and (\ref{back2}) give
\ba
A_0' &=& \frac{q_E}{r^2 {\cal L}_{,F}}\,,
\label{Leq1}\\
{\cal L} &=& r^{-2} \left[ q_E A_0'+\Mpl^2 
(rf'+f-1) \right]\,,
\label{Leq2}
\ea
where $q_E$ is a constant corresponding to an electric charge. 
Taking the $r$-derivative of Eq.~(\ref{Leq2}) and combining 
it with Eq.~(\ref{Leq1}) to eliminate 
${\cal L}_{,F}$, we obtain
\be
2q_E r^4 A_0'^2+\Mpl^2 \left( 2f-2-r^2 f'' \right) 
r^4 A_0'+2q_E q_M^2=0\,,
\label{back4}
\ee
which algebraically determines $A_0'$ in terms of 
$f$ and its second radial derivative.

We are interested in nonsingular BHs with regular centers. 
To avoid the singularities of Ricci scalar $R$, Ricci squared $R_{\mu \nu}R^{\mu \nu}$, Riemann squared $R_{\mu \nu \rho \sigma}R^{\mu \nu \rho \sigma}$ at $r=0$, we require that $f$ is expanded around $r=0$ as \cite{Frolov:2016pav}
\be
f(r)=1+\sum_{n=2}^{\infty} f_n r^n\,,
\label{fexpan}
\ee
where $f_n$'s are constants. 
The deviation of $f(0)$ from 1 results in conical singularities. 
Any power of $n$ smaller than 2 leads to curvature singularities.
Substituting Eq.~(\ref{fexpan}) and its second radial derivative into Eq.~(\ref{back4}), two branches of $A_0'$ have the leading-order terms $\pm \sqrt{-(q_E q_M)^2}/(q_E r^2)$. 
This means that there exist real solutions to $A_0'$ only if
\be
q_E q_M=0\,.
\label{qEM}
\ee
Thus, the presence of dyon BHs with $q_E \neq 0$ and $q_M \neq 0$ is forbidden from the regularity of $f$ at $r=0$. From Eq.~(\ref{qEM}), either $q_E$ or $q_M$ must be 0. 
This non-existence of regular dyon BH solutions breaks the electromagnetic duality present in linear electrodynamics, 
where the property of BHs is determined by their mass and 
total charge $q_T=\sqrt{q_E^2+q_M^2}$ (see e.g.,\ 
\cite{MISNER1957525,DeFelice:2023rra}).

\subsection{Purely electric BHs}

For $q_E \neq 0$ and $q_M=0$, the nonvanishing solution to $A_0'$ follows from Eq.~(\ref{back4}), such that 
\be
A_0'=\frac{\Mpl^2 (r^2 f''-2f+2)}{2q_E}\,.
\label{A0dE}
\ee
Using the regular metric (\ref{fexpan}) 
around $r=0$, we have 
\be
A_0'=\frac{2\Mpl^2 f_3}{q_E}\,r^3+
\frac{5\Mpl^2 f_4}{q_E}r^4+{\cal O}(r^5)\,,
\ee
which approaches 0 as $r \to 0$.
Substituting Eq.~\eqref{A0dE} into Eqs.~(\ref{Fexp}) and (\ref{Leq2}), we obtain
\ba
F &=&\frac{\Mpl^4 (r^2 f''-2f+2)^2}{8q_E^2}\,,
\label{F1}\\
{\cal L} &=& \frac{\Mpl^2}{2} \left( f''
+\frac{2f'}{r} \right)\,.
\label{L1}
\ea
Around $r=0$, these behave as $F=2\Mpl^4 f_3^2r^6/q_E^2+{\cal O}(r^7)$ and ${\cal L}=3\Mpl^2 (f_2+2f_3 r)+{\cal O}(r^2)$, 
which are both finite.
 
The nonsingular BH proposed by Ayon-Beato and 
Garcia \cite{Ayon-Beato:1998hmi} is characterized 
by the metric components
\be
f=h=1-\frac{2M r^2}{(r^2+r_0^2)^{3/2}}
+\frac{r_0^2 r^2}{(r^2+r_0^2)^2}\,,
\label{Ayon-Beato}
\ee
where $M$ and $r_0$ are constants. 
At large distances, Eq.~(\ref{Ayon-Beato}) 
approaches the Reissner-Nordstr\"om (RN) metric 
components $f=h=1-2M/r+q_E^2/(2\Mpl^2 r^2)$, 
with the correspondence $q_E=\sqrt{2}\Mpl r_0$. 
Around $r=0$, the metric (\ref{Ayon-Beato}) 
is related to the coefficients in Eq.~(\ref{fexpan}) as $f_2=-(2M-r_0)/r_0^3$, $f_3=0$, and $f_4=(3M-2r_0)/r_0^5$. 
So long as $f_2<0$, i.e., $r_0<2M$, the central region 
of BHs is approximately described by the de Sitter spacetime, 
which generates pressure against gravity.

For a given regular metric $f$, we know both $F$ and ${\cal L}$ as functions of $r$ from Eqs.~(\ref{F1}) and (\ref{L1}). 
In this case, we can also express ${\cal L}$ as a function of $F$, provided that $r$ is explicitly written in terms of $F$.

\subsection{Purely magnetic BHs}

For $q_M \neq 0$ and $q_E=0$, 
Eqs.~(\ref{Leq1}) and (\ref{Leq2}) give 
\ba
A_0'&=& 0\,,\\
{\cal L} &=& \Mpl^2 r^{-2} (rf'+f-1)\,,
\label{LM}
\ea
with $F=-q_M^2/(2r^4)$. 
Using the expansion (\ref{fexpan}), 
the Lagrangian is regular as ${\cal L}=\Mpl^2(3f_2+4f_3r)+\mathcal{O}(r^2)$ around $r=0$.
For a given $f(r)$, we can explicitly express
${\cal L}$ as a function of $F$ by using Eq.~(\ref{LM}). 
The nonsingular BH proposed by Dymnikova \cite{Dymnikova:2004zc} 
corresponds to the metric components
\be
f=h=1-\frac{4M}{\pi r} \left[ \arctan 
\left( \frac{r}{r_0} 
\right)-\frac{r_0 r}{r^2+r_0^2} \right]\,,
\label{Dymnikova}
\ee
where $r_0=\pi q_M^2/(16 \Mpl^2 M)$ to recover 
the magnetic RN solution 
$f=1-2M/r+q_M^2/(2\Mpl^2 r^2)$ at large distances. 
In this case, the Lagrangian is known as 
\be
{\cal L}=-\frac{q_M^2}{2(r^2+r_0^2)^2}
=-\frac{q_M^2}
{(\sqrt{2}r_0^2+\sqrt{-q_M^2/F})^2}\,.
\label{DymnikovaL}
\ee
This recovers the standard Maxwell Lagrangian ${\cal L}=F$ 
as $F \to -0$ (i.e., in the limit $r \to \infty$).

\section{Angular Laplacian instabilities of nonsingular BHs} 
\label{stasec}

To study the linear stability of electric and 
magnetic BHs, we consider metric and vector-field 
perturbations on the SSS background 
(\ref{metric}) \cite{Regge:1957td,Zerilli:1970se,Moncrief:1974ng}.
For the components of metric perturbations 
$h_{\mu \nu}$, we choose
\ba
\hspace{-0.1cm}
& &
h_{tt}=f(r) H_0 (t,r) Y_{l}(\theta), \quad 
h_{tr}=H_1 (t,r) Y_{l}(\theta),\quad 
h_{t \theta}=0, 
\nonumber \\
\hspace{-0.1cm}
& &
h_{t \varphi}=-Q(t,r) (\sin \theta) 
Y_{l, \theta} (\theta),
\;\; 
h_{rr}=f^{-1}(r) H_2(t,r) Y_{l}(\theta),\nonumber \\
\hspace{-0.1cm}
& &
h_{r \theta}=h_1 (t,r)Y_{l, \theta}(\theta),
\quad
h_{r \varphi}=-W(t,r) (\sin \theta) Y_{l,\theta} (\theta), 
\nonumber \\
\hspace{-0.1cm}
& &
h_{\theta \theta}=0,\quad 
h_{\varphi \varphi}=0,\quad
h_{\theta \varphi}=0,
\label{hcom}
\ea
where $Y_{l}(\theta)$ is the $m=0$ component of 
spherical harmonics $Y_{lm}(\theta, \varphi)$. 
On the SSS background, we can focus on the axisymmetric 
modes ($m=0$) without loss of generality.
The covector-field perturbation $\delta A_{\mu}$ 
has the following components
\ba
& &
\delta A_t=\delta A_0 (t,r) Y_{l}(\theta),\qquad 
\delta A_r=\delta A_1 (t,r) Y_{l}(\theta),\qquad \nonumber \\
& &
\delta A_\theta=0,\qquad 
\delta A_{\varphi}=-\delta A(t,r) (\sin \theta) 
Y_{l,\theta}(\theta)\,,
\label{perma}
\ea
where the choice $\delta A_\theta=0$ is an outcome of the presence of $U(1)$ gauge symmetry.\footnote{Since there is $U(1)$ gauge 
invariance under the transformation, $A_\mu\to A_\mu+\partial_\mu\chi_{\rm gauge}$, we can eliminate 
the even-parity mode $\delta A_\theta=\delta A_2(t,r) Y_{l,\theta}(\theta)$ by choosing the perturbed gauge field $\delta\chi_{\rm gauge}=-\delta A_2(t,r) Y_l(\theta)$ and making the field redefinitions $\delta A_0^{\rm new}= \delta A_0 - \dot{\delta A}_2$, $\delta A_1^{\rm new}=\delta A_1-\delta A_2'$.} 
We note that the gauge choice (\ref{hcom}) completely fixes the residual gauge degrees of freedom under the infinitesimal transformation $x^{\mu} \to x^{\mu}+\xi^{\mu}$.

The three perturbations $Q$, $W$, $\delta A$ belong to those in the odd-parity sector, while the six perturbations $H_0$, $H_1$, $H_2$, $h_1$, $\delta A_0$, $\delta A_1$ correspond to those in the even-parity sector.
We focus on the multiple modes $l \geq 2$ and expand the action (\ref{action}) up to second order in perturbed fields. 
The total quadratic-order action can be expressed as ${\cal S}^{(2)}=\int \rd t \rd r\,({\cal L}_1+{\cal L}_2)$, where ${\cal L}_1$ and ${\cal L}_2$ are given in Appendix~A. 
We introduce the following Lagrange multipliers 
\ba
\chi &=& 
r\dot{W}-rQ'+2Q-\frac{2{\cal L}_{,F}rA_0'}
{\Mpl^2}\delta A\,, \\
V &=& 
\delta A_0'-\dot{\delta A_1}+\frac{A_0'}{2} 
(H_0-H_2)\,,
\ea
where a dot represents the derivative with respect to $t$.
The dynamical fields $\chi$ and $V$ correspond to the odd-parity gravitational perturbation and the even-parity electromagnetic perturbation, respectively.
We also have the odd-parity electromagnetic mode $\delta A$ and the even-parity gravitational mode $\psi$ defined by 
\ba
\psi=r H_2- L h_1\,,\quad {\rm where} 
\quad L=l(l+1)\,.
\ea
Following the procedure explained in Appendix A, 
the second-order action, after the elimination of 
all nondynamical perturbations and the integration 
by parts, is expressed in the form 
\be
\tilde{{\cal S}}^{(2)}=\int \rd t \rd r
\left( \dot{\vec{\Psi}}^t{\bm K}\dot{\vec{\Psi}}
+\vec{\Psi}'^{t}{\bm G}\vec{\Psi}'
+\vec{\Psi}^{t}{\bm M}\vec{\Psi}
+\vec{\Psi}'^{t}{\bm S}\vec{\Psi} \right)
\,,\label{secondLag}
\ee
where ${\bm K}, {\bm G}, {\bm M}$ are $4 \times 4$ symmetric matrices with components like $K_{11}$, ${\bm S}$ is a $4 \times 4$ antisymmetric matrix, and 
\be
\vec{\Psi}^t= \left( \chi, \delta A, \psi, V \right)\,.
\ee
In the eikonal limit ($l \gg 1$), we will derive 
the linear stability conditions for 
electric and magnetic BHs. 
Unlike past related works \cite{Moreno:2002gg, Nomura:2020tpc}, 
our results can be applied to the stability for 
both $f>0$ and $f<0$.

\subsection{Purely electric BHs}

For $q_E \neq 0$ and $q_M=0$, the dynamical system of perturbations is decomposed into the odd-parity sector with $\vec{\Psi}_{\rm A}^t =( \chi, \delta A)$ and the even-parity sector with $\vec{\Psi}_{\rm B}^t
=( \psi, V)$. When $f>0$, the positivities of 
${\bm K}_{\rm A}$ and ${\bm K}_{\rm B}$, 
which are the $2 \times 2$ kinetic matrices of ${\bm K}$ associated with $\vec{\Psi}_{\rm A}^t$ and $\vec{\Psi}_{\rm B}^t$ respectively, determine 
the no-ghost conditions of four dynamical 
perturbations. So long as 
\be
{\cal L}_{,F}>0\,,
\label{NG}
\ee
both ${\bm K}_{\rm A}$ and ${\bm K}_{\rm B}$ 
are positive definite. 
For $f<0$, the no-ghost conditions are determined 
by the positivities of matrices ${\bm G}_{\rm A}$ 
and ${\bm G}_{\rm B}$ associated with 
$\vec{\Psi}_{\rm A}^t$ and 
$\vec{\Psi}_{\rm B}^t$ respectively.
They are satisfied with the inequality (\ref{NG}).

For $f>0$, the radial propagation speeds $c_r$ measured 
by a proper time $\tau=\int f\,\rd t$
are known by substituting the WKB-form solutions $\vec{\Psi}^t=\vec{\Psi}_0^t e^{-i (\omega t-kr)}$ into 
their perturbation equations, where 
$\vec{\Psi}_0^t$ is a constant vector composed of $(\vec{\Psi}_0^t)_{\rm A}$ and 
$(\vec{\Psi}_0^t)_{\rm B}$.
This leads to the algebraic equations 
${\bm U}_{\rm A} (\vec{\Psi}_0)_{\rm A}=0$ and 
${\bm U}_{\rm B} (\vec{\Psi}_0)_{\rm B}=0$, 
where ${\bm U}_{\rm A}$ and ${\bm U}_{\rm B}$ 
are $2 \times 2$ matrices. 
The existence of nonvanishing solutions to 
$(\vec{\Psi}_0)_{\rm A}$ and $(\vec{\Psi}_0)_{\rm B}$ requires that ${\rm det}\,{\bm U}_{\rm A}=0$ 
and ${\rm det}\,{\bm U}_{\rm B}=0$. 
Taking the large $\omega$ and $k$ limits, 
both equations lead to $(\omega^2-k^2 f^2)^2=0$. 
Substituting $\omega=k f c_r$ into this relation, we find
\be
c_r^2=1\,,\qquad {\rm for}~~~{\rm all}\quad 
\vec{\Psi}^t= \left( \chi, \delta A, \psi, V \right)\,.
\label{cr}
\ee
When $f<0$, we exploit the WKB solution in the form $\vec{\Psi}^t=\vec{\Psi}_0^t e^{-i (\omega r-kt)}$. 
This results in the dispersion relation 
$(\omega^2 f^2-k^2)^2=0$ for both 
$\vec{\Psi}_{\rm A}^t$ and $\vec{\Psi}_{\rm B}^t$. 
Then, after the substitution of $\omega=kc_r/(-f)$, 
we obtain the same squared radial propagation speeds 
as those in Eq.~(\ref{cr}).

For $f>0$, the angular propagation speeds $c_\Omega$ 
are derived by taking the large $\omega$ 
and $l$ limits in ${\rm det}\,{\bm U}_{\rm A}=0$ and 
${\rm det}\,{\bm U}_{\rm B}=0$.
{}From ${\rm det}\,{\bm U}_{\rm A}=0$, we obtain the dispersion relation $(r^2 \omega^2-L f)^2=0$. Substituting 
$\omega=c_{\Omega} l \sqrt{f}/r$ into this relation and 
taking the limit $l \gg 1$, we find
\be
c_{\Omega}^2=1\,,\qquad {\rm for} \quad 
\vec{\Psi}_{\rm A}^t= \left( \chi, \delta A \right)\,.
\label{cO1}
\ee
The two solutions following from 
${\rm det}\,{\bm U}_{\rm B}=0$ are 
\ba
& &
c_{\Omega}^2=1\,,\qquad {\rm for}~~\psi\,,\label{cO2}\\
& &
c_{\Omega}^2=c_E^2 \equiv \frac{{\cal L}_{,F}}
{{\cal L}_{,F}+2F{\cal L}_{,FF}}\,,
\qquad {\rm for}~~V\,.
\label{cO3}
\ea
Since $M_{44}/K_{44}=-c_E^2(l^2f/r^2)$ for $l \gg 1$, we can identity $c_E^2$ as the squared angular propagation speed of $V$. 
When $f<0$, using the WKB-form solution $\vec{\Psi}^t=\vec{\Psi}_0^t e^{-i (\omega r-kt)}$ with $\omega=c_{\Omega} l/(\sqrt{-f}\,r)$ results in the 
same values of $c_\Omega^2$ as those given in 
Eqs.~(\ref{cO1})-(\ref{cO3}).

\subsection{Purely magnetic BHs}

For $q_M \neq 0$ and $q_E=0$, the system is separated 
into two sectors: type (C) with 
$\vec{\Psi}_{\rm C}^t=( \chi, V)$ and type (D) with $\vec{\Psi}_{\rm D}^t=( \delta A, \psi)$ \cite{Nomura:2020tpc,Chen:2024hkm}.
When $f>0$, the positivities of $2 \times 2$ kinetic matrices 
${\bm K}_{\rm C}$ and ${\bm K}_{\rm D}$ in each sector are ensured under the condition (\ref{NG}). 
This is also the case for $f<0$, where the positivities of matrices ${\bm G}_{\rm C}$ and ${\bm G}_{\rm D}$ determine the no-ghost conditions. 

Using the WKB-form solution 
$\vec{\Psi}^t=\vec{\Psi}_0^t e^{-i (\omega t-kr)}$ for $f>0$, we obtain the two algebraic equations 
${\bm U}_{\rm C} (\vec{\Psi}_0)_{\rm C}=0$ and 
${\bm U}_{\rm D} (\vec{\Psi}_0)_{\rm D}=0$ 
in type C and D sectors, respectively.
Taking the large $\omega$ and $k$ limits for 
${\rm det}\,{\bm U}_{\rm C}=0$ and 
${\rm det}\,{\bm U}_{\rm D}=0$, we find that 
all four dynamical perturbations have the luminal squared radial propagation speeds $c_r^2=1$. 
This property also holds for $f<0$. 

For the sector (C) with $f>0$, taking the large $\omega$ and $l$ limits for ${\rm det}\,{\bm U}_{\rm C}=0$ leads to the squared angular propagation speeds
\be
c_{\Omega}^2=1\,,\qquad {\rm for} \quad 
\vec{\Psi}_{\rm C}^t= \left( \chi, V \right)\,.
\label{cOM}
\ee
From the other equation ${\rm det}~{\bm U}_{\rm D}=0$, 
we obtain
\ba
& &
c_{\Omega}^2=c_M^2 \equiv 
\frac{{\cal L}_{,F}+2F{\cal L}_{,FF}}{{\cal L}_{,F}}
\,,\qquad {\rm for}~~\delta A\,,\label{cO2M}\\
& &
c_{\Omega}^2=1\,,
\qquad {\rm for}~~\psi\,.
\label{cO3M}
\ea
Since $M_{22}/K_{22}=-c_M^2 l^2 f/r^2$ for $l \gg 1$, we can identity $c_M^2$ as the squared angular 
propagation speed of $\delta A$. 
Unlike the electric BH, the odd-parity electromagnetic perturbation $\delta A$ has a nontrivial propagation speed different from 1. 
Again, the results (\ref{cOM})-(\ref{cO3M}) are valid for $f<0$.

\section{Instability of nonsingular BHs} 
\label{inssec}

For the electric BH, we compute Eq.~(\ref{cO3}) by 
differentiating Eq.~(\ref{Leq1}) and using 
the relation $F=A_0'^2/2$. 
This gives ${\cal L}_{,FF}=-q_E (rA_0''+2A_0')/(r^3 A_0'' A_0'^3)$ and hence $c_E^2=-r A_0''/(2A_0')$. 
By using Eq.~(\ref{A0dE}), we obtain
\be
c_E^2=c_f^2 \equiv -\frac{r(r^2 f'''+2rf''-2f')}
{2(r^2 f''-2f+2)}\,,
\label{cf}
\ee
which depends on $f$ and its $r$ derivatives alone.

For the magnetic BH, we take the $F$ derivative of 
Eq.~(\ref{LM}) and exploit the relation $F=-q_M^2/(2r^4)$. Then, we find that $c_M^2$ in Eq.~(\ref{cO2M}) 
reduces to $c_f^2$ in Eq.~(\ref{cf}).
Thus, for a given metric function $f(r)$, 
the squared angular propagation speeds 
$c_E^2$ and $c_M^2$ can be expressed 
in a unified manner. 
We have $c_f^2=1$ at any distance $r$ for the RN 
metric $f=1-2M/r+q^2/(2\Mpl^2 r^2)$, 
but this property does not hold for nonsingular BHs.

Let us consider the nonsingular BH with the expansion (\ref{fexpan}) of $f$ around $r=0$. 
Since $f>0$ in this regime, the $t$ and $r$ coordinates 
play the timelike and spacelike roles, respectively.
The expansion of $c_f^2$ leads to
\be
c_f^2=-\frac{3}{2}-\frac{5f_4}{4f_3}r
+\frac{25f_4^2-36f_3 f_5}{8f_3^2}r^2
+{\cal O}(r^3)\,,
\label{cf1}
\ee
which is valid for $f_3 \neq 0$. 
Nonsingular BHs like (\ref{Ayon-Beato}) and (\ref{Dymnikova}) correspond to $f_3=0$, in which case we have
\be
c_f^2=-2-\frac{9f_5}{10f_4}r
+\frac{81f_5^2-140f_4 f_6}{50 f_4^2}r^2
+{\cal O}(r^3)\,.
\label{cf2}
\ee
Thus, in both cases, the leading-order terms 
of $c_f^2$ are negative.
For the metric function 
$f=1+f_n r^n+\mathcal{O}(r^{n+1})$, 
we have $c_f^2=-n/2+{\cal O}(r)$ and 
hence $c_f^2 \le -1$ for $n \geq 2$.

We study the behavior of dynamical perturbations 
$V$ and $\psi$ around $r=0$ for the electric BH.
Expressing those fields as $V=\tilde{V}(t) e^{ikr}$ and $\psi=\tilde{\psi}(t) e^{ikr}$ for $f>0$ and taking the limits 
$l \gg 1$ and $l\gg kr$, the time-dependent parts 
approximately obey the differential equations 
\begin{align}
\ddot{\tilde{V}}&+c_f^2\,\frac{ f l^2}{r^2} \tilde{V}
=\frac{f^{2} \Omega_{f} c_{f}^{4} \Mpl^2}{r^{3} q_{E}} \tilde{\psi}\,,
\label{per1}\\
\ddot{\tilde{\psi}}&+\frac{fl^2}{r^2} \tilde{\psi}
=\frac{l^{2} q_{E}}{r c_{f}^{2} \Mpl^2}\tilde{V}\,,
\label{per2}
\end{align}
where $\Omega_f \equiv r^2f''-2f+2$. 
If we use the expansion $f=1+f_n r_n+{\cal O}(r^{n+1})$ 
around $r=0$, 
we have that 
$\Omega_f=(n^2-n-2)\,f_nr^n + \mathcal{O}(r^{n+1})$. 
It is possible to close Eqs.~\eqref{per1} and \eqref{per2} 
for one single variable, say $\tilde{\psi}$, finding
\begin{equation}
\ddddot{\tilde\psi}+\frac{(1+c_f^2)\,f l^{2}}{r^{2}}
\ddot{\tilde\psi}+\frac{c_f^2 f^2 l^4}{r^{4}}\tilde\psi 
\simeq 0\,,\label{eq:ang_prop_psi}
\end{equation}
where we have neglected the term 
$-(f^2 c_f^2\Omega_f l^2/r^4)\tilde\psi$.
Assuming the solution to Eq.~(\ref{eq:ang_prop_psi}) 
in the form $\tilde \psi(t) \propto e^{-i\omega t}$, 
we obtain
\begin{equation}
\omega^2=c_f^2\,\frac{f l^2}{r^2}\,,\qquad
\omega^2=\frac{f l^2}{r^2}\,.
\label{eq:class_inst}
\end{equation}
{}Since $c_f^2<0$ in a non-empty set centered around $r=0$, 
there is always a growing-mode solution 
($\omega^2=c_f^2 f l^2/r^2$)
besides a stable one ($\omega^2=f l^2/r^2$). 
However, the presence of the former is enough 
to make the nonsingular BH unstable. 
We note that $\tilde{V}$ obeys the same form of 
a fourth-order differential equation 
as Eq.~(\ref{eq:ang_prop_psi}), 
so that the two dynamical perturbations 
$\psi$ and $V$ in the even-parity sector (B) are 
subject to exponential growth.

The enhancement of $\psi$ and $V$ works as 
a backreaction to the background BH solution.
Then, the background metric is no longer maintained 
as the steady forms like (\ref{Ayon-Beato}) 
and (\ref{Dymnikova}). 
For the magnetic BH, the same exponential 
growth of dynamical perturbations occurs for 
$\psi$ and $\delta A$ in the sector (D).
Such instability is generic for all nonsingular BHs 
constructed in the framework of NED--including those of 
Bardeen with metric $f=1-2M r^2/(r^2+r_0^2)^{3/2}$ \cite{Bardeen:1968}
and Hayward with metric $f=1-2M r^2/(r^3+2M r_0^2)$ \cite{Hayward:2005gi}.

A typical time scale of instability arising from 
the negative value of $c_f^2$ around $r=0$ is estimated as 
\be
t_{\rm ins} \simeq
\frac{r}{\sqrt{-c_f^2} l}\,.
\label{tins}
\ee
We recall that $-c_f^2$ is of order $c^2$, where we restored 
the speed of light $c$.
Since the distance $r$ associated with Laplacian instability 
is less than the outer horizon radius $r_h$, $t_{\rm ins}$ 
is much shorter than $r_h/c$ for $l \gg 1$. 
If we consider a BH with $r_h=10$~km, 
we have $t_{\rm ins} \lesssim 10^{-5}/l$~sec.

\begin{figure}[ht]
\begin{center}
\includegraphics[height=3.0in,width=3.0in]{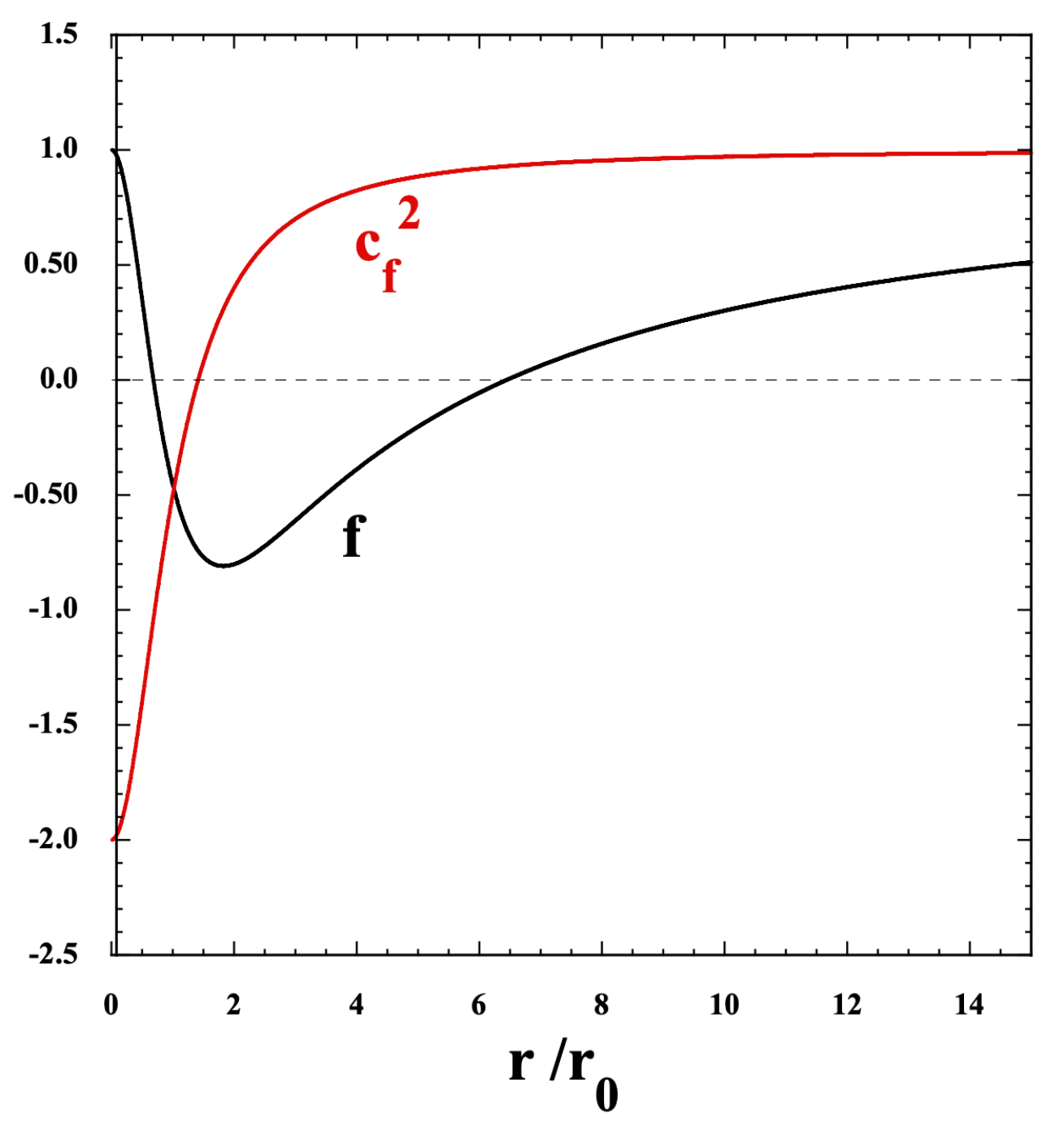}
\end{center}\vspace{-0.5cm}
\caption{\label{fig1}
We plot $c_f^2$ and $f$ versus $r/r_0$ 
for the metric (\ref{Dymnikova}) with $M=4r_0$. 
At the distance $r<\sqrt{2}r_0$, we have $c_f^2<0$. 
For the metric \eqref{Ayon-Beato}, $c_f^2$ exhibits 
similar behavior around $r=0$.}
\end{figure}

The above results show that nonsingular BHs in NED 
are always plagued by angular Laplacian instability around $r=0$.
For example, the BH solution (\ref{Dymnikova}) has the 
following squared angular propagation speed
\be
c_f^2=\frac{r^2-2r_0^2}{r^2+r_0^2}\,.
\label{cfe}
\ee
As we estimated in Eq.~(\ref{cf2}), we have 
$c_f^2=-2$ at $r=0$.
While $c_f^2$ approaches 1 as $r \to \infty$, 
$c_f^2$ is negative in the region $r<\sqrt{2}r_0$. 

In Fig.~\ref{fig1}, we plot $c_f^2$ and $f$  for the metric (\ref{Dymnikova}) with $M=4r_0$, in which case there are two horizons at $r_1=0.69r_0$ and $r_2=6.44r_0$.
Since the expression (\ref{cfe}) is valid at any 
distance $r \geq 0$,
there is angular Laplacian instability for $r<\sqrt{2}r_0$ 
(including the region $r_1 \le r<\sqrt{2}r_0$ with $f<0$). 
The crucial point is that nonsingular BHs always have a finite range of $r$ where $f$ is expanded as Eq.~(\ref{fexpan}) around $r=0$, in which regime $c_f^2$ is always negative. 
In Appendix B, we will confirm that the angular Laplacian 
instability is robust 
irrespective of the presence/absence of ghosts 
and the rescaling of dynamical perturbations.

\section{Conclusions} 
\label{consec}

We have shown that nonsingular BHs in NED are inevitably subject to angular Laplacian instability around $r=0$. 
This result holds for both electric and magnetic BHs, as the form (\ref{cf}) of $c_f^2$ is universal to both cases. 
The Laplacian instability we found is a physical one, in that the 
even-parity gravitational perturbation $\psi$ is subject to 
exponential growth through the angular instability of 
vector-field perturbations ($V$ for the electric BH and 
$\delta A$ for the magnetic BH). 
The backreaction of enhanced perturbations to the background
would not keep the regular metrics 
like (\ref{Ayon-Beato}) and (\ref{Dymnikova}) as they are.

Our no-go result for the absence of stable static nonsingular BHs 
is valid for NED, but this is not the case for more general theories. For example, it is of interest to study what happens by incorporating an additional scalar field $\phi$ as the Lagrangian ${\cal L}(\phi, X, F)$ \cite{Heisenberg:2018acv,Kase:2023kvq,Pereira:2024rtv}, where $X$ is a scalar kinetic term. If such theories with dynamical degrees of freedom still lead to the instability of regular BHs, 
nonlocal versions of the ultraviolet completion of gravity 
such as those proposed in 
Refs.~\cite{Modesto:2011kw, Biswas:2011ar, Modesto:2014lga, Tomboulis:2015gfa} may be the clue to the construction of stable nonsingular BHs. 

\section*{Acknowledgements}

We thank Valeri Frolov for useful discussions.
The work of ADF was supported by the Japan Society for the Promotion of Science Grants-in-Aid for Scientific Research No.~20K03969.
ST was supported by the Grant-in-Aid for Scientific Research Fund of the JSPS No.~22K03642 and Waseda University Special Research Project No.~2024C-474. 

\section*{Appendix~A:~Second-order perturbed action}
\renewcommand{\theequation}{A.\arabic{equation}} 
\setcounter{equation}{0}

The second-order action of perturbations, which is obtained 
after the integration with respect to $\theta$ and $\varphi$, 
can be written 
in the form ${\cal S}^{(2)}=\int \rd t \rd r\,
({\cal L}_1+{\cal L}_2)$, where
\begin{widetext}
\ba
{\cal L}_1 &=&
a_0 H_0^2 
+ H_0 \left[ a_1 H_2' + L a_2 h_1' 
+(a_3+L a_4)H_2 + L a_5 h_1+L a_6 \delta A \right]
+Lb_1 H_1^2+H_1 ( b_2 \dot{H}_2+L b_3 \dot{h}_1) \nonumber \\
& &+c_0 H_2^2 +L H_2 ( c_1 h_1+c_2 \delta A)
+L (d_0 \dot{h}_1^2+ d_1 h_1^2)
+L h_1 (d_2 \delta A_0+d_3 \delta A') \nonumber \\
& &
+s_1 ( \delta A_0'-\dot{\delta A_1} )^2
+(s_2 H_0 +s_3 H_2+L s_4 \delta A)( \delta A_0'-\dot{\delta A_1}) 
+L ( s_5 \delta A_0^2 +s_6 \delta A_1^2)\,,
\label{LagA}\\
{\cal L}_2 &=& L [ p_1 (r\dot{W}-r Q'+2Q)^2+p_2 \delta A 
(r\dot{W}-r Q'+2Q)+p_3 \dot{\delta A^2}+p_4 \delta A'^2
+L p_5 \delta A^2+(L p_6+p_7)W^2 \nonumber \\
& &~~+(L p_8+p_9)Q^2+p_{10} Q \delta A_0
+p_{11}Q h_1+p_{12} W \delta A_1]\,,
\label{LagB}
\ea
where we used the condition $h=f$, and
\ba
& &
a_0=\frac{r^2}{8}A_0'^2 ({\cal L}_{,F}+A_0'^2 {\cal L}_{,FF}),\qquad
a_1=-\frac{\Mpl^2 rf}{2},\qquad
a_2=\frac{\Mpl^2 f}{2},\qquad
a_3=-\frac{\Mpl^2}{2}-\frac{r^2}{4}(2{\cal L}
-A_0'^2 {\cal L}_{,F}+A_0'^4 {\cal L}_{,FF}), \nonumber \\
& &
a_4=-\frac{\Mpl^2}{4},\qquad 
a_5=\frac{\Mpl^2 (f+1)+r^2({\cal L}-A_0'^2 {\cal L}_{,F})}{4r},
\qquad
a_6=\frac{q_M}{2r^2}({\cal L}_{,F}-A_0'^2 {\cal L}_{,FF}),
\qquad 
b_1=-a_4,\nonumber \\
& &
b_2=\Mpl^2 r,\qquad b_3=2a_4\,,\qquad c_0=-\frac{a_3}{2},
\qquad 
c_1=-a_5,\qquad c_2=-a_6,\qquad d_0=-a_4,\qquad 
d_1=\frac{f}{2r^4}(\Mpl^2 r^2-q_M^2{\cal L}_{,F}), \nonumber \\
& & d_2=-A_0'{\cal L}_{,F},\qquad 
d_3=-\frac{q_M f{\cal L}_{,F}}{r^2},\qquad 
s_1=\frac{r^2}{2}({\cal L}_{,F}+A_0'^2 {\cal L}_{,FF}),\qquad 
s_2=A_0' s_1,\qquad s_3=-A_0' s_1,\nonumber \\
& &
s_4=-\frac{q_M A_0'{\cal L}_{,FF}}{r^2},\qquad
s_5=\frac{{\cal L}_{,F}}{2f},\qquad 
s_6=-\frac{f{\cal L}_{,F}}{2},\label{coeff1}\\
& & p_1=\frac{\Mpl^2}{4r^2},\qquad p_2=\frac{d_2}{r},\qquad p_3=s_5,\qquad 
p_4=s_6,\qquad p_5=\frac{q_M^2{\cal L}_{,FF}-r^4{\cal L}_{,F}}{2r^6},
\qquad p_6=-\frac{\Mpl^2 f}{4r^2},\nonumber \\
& & 
p_7=d_1,\qquad p_8=\frac{\Mpl^2}{4r^2f},\qquad 
p_9=-\frac{d_1}{f^2},\qquad p_{10}=-\frac{d_3}{f^2},\qquad 
p_{11}=-A_0'f p_{10},\qquad p_{12}=-f^2 p_{10}\,,\label{coeff2}
\ea
\end{widetext}
where $s_4$ vanishes for both the electric BH ($q_M=0$) 
and the magnetic BH ($A_0'=0$). 
The second-order Lagrangians (\ref{LagA}) and (\ref{LagB}) 
with the coefficients (\ref{coeff1}) and (\ref{coeff2}) 
are valid both for $f>0$ and $f<0$.

We incorporate the dynamical fields 
$V$ and $\chi$ as the forms of Lagrange multipliers 
\ba
\hspace{-0.2cm}
\tilde{{\cal L}}_1 &=&
{\cal L}_1-s_1 
\left[ \delta A_0'-\dot{\delta A_1}
+\frac{A_0'}{2} (H_0-H_2)-V \right]^2\,,\\
\hspace{-0.2cm}
\tilde{{\cal L}}_2 &=&
{\cal L}_2-L p_1 
\left[ r\dot{W}-rQ'+2Q-\frac{2{\cal L}_{,F}rA_0'}
{\Mpl^2}\delta A-\chi \right]^2\,.\nonumber\\
\ea
Then, we consider the action 
$\tilde{{\cal S}}^{(2)}=\int \rd t \rd r\,
(\tilde{{\cal L}}_1+\tilde{{\cal L}}_2)$ equivalent 
to ${\cal S}^{(2)}$. 
We also introduce the dynamical perturbation 
$\psi=rH_2-L h_1$ and express $H_2$ 
in terms of $\psi$ and $h_1$. Since $H_0^2$ vanishes 
in $\tilde{{\cal L}}_1$, the variation of   
$\tilde{{\cal S}}^{(2)}$ with respect to $H_0$ 
allows one to express $h_1$ in terms of 
the other fields.
After deriving the perturbation equations of motion 
for $H_1$, $\delta A_0$, $\delta A_1$, $Q$, and $W$, 
we can eliminate these fields from $\tilde{{\cal S}}^{(2)}$.
After the integration by parts, we finally obtain 
the second-order action of the form (\ref{secondLag}) 
containing four dynamical perturbations 
$\chi$, $\delta A$, $\psi$, $V$, and 
their $t, r$ derivatives.

\section*{Appendix~B:~No-ghost conditions}
\setcounter{equation}{0}

Let us discuss the no-ghost conditions
in more detail. For the electric BH, we have 
${\cal L}_{,F}=q_E/(r^2 A_0')=2q_E^2/(\Mpl^2 r^2 \Omega_f)$ 
from Eqs.~(\ref{Leq1}) and (\ref{A0dE}). 
For the magnetic BH, we obtain 
${\cal L}_{,F}=\Mpl^2 r^2 \Omega_f/(2q_M^2)$ from 
Eq.~(\ref{LM}). 
Then, in both cases, the no-ghost condition 
(\ref{NG}) is equivalent to 
\begin{equation}
\Omega_f=r^2 f''-2f+2 > 0\,.
\label{eq:nogh}
\end{equation}
Using the expansion (\ref{fexpan}) around $r=0$, 
this inequality translates to 
$\Omega_f= 4 f_3 r^3+ 10 f_4 r^4+{\cal O}(r^5)>0$, 
which is always satisfied if $f_3>0$ (and if $f_4>0$ 
for the BH solution with $f_3=0$).

For the electric BH in the range $f>0$, the second-order action of even-parity perturbations contains kinetic terms of 
$V$ and $\psi$, as
\be
\tilde{{\cal S}}^{(2)}=\int {\rm d}t\,{\rm d}r 
\left( \frac{q_E^2 r^2}{l^2f \Omega_f \Mpl^2c_f^4}\,
\dot{V}^2+\frac{\Mpl^2 f}{l^4}\dot{\psi}^2 \cdots \right)\,,
\label{actionkin}
\ee
Thus, in the limit that $r \to 0$, there is no strong coupling 
associated with the vanishing kinetic terms.
Under the no-ghost condition $\Omega_f>0$ together 
with the regular condition $f>0$ around $r=0$, 
the coefficients of 
$\dot{V}^2$ and $\dot{\psi}^2$ are both positive. 
One can perform the field definitions for
$V$ and $\psi$ to make the kinetic terms 
in Eq.~(\ref{actionkin}) canonical. 
However, this does not modify the squared angular 
propagation speed $c_f^2$. 
Indeed, Eq.~\eqref{eq:ang_prop_psi} shows the invariance 
under the field redefinition $\tilde{\psi} 
\to {\cal F}(r,l)\tilde{\psi}$, where ${\cal F}$ 
depends on $r$ and $l$. 
For the magnetic BH, the same property 
for the invariance of $c_f^2$ also holds under the 
redefinition of $\delta A$ and $\psi$. 
Therefore, the angular instability around $r=0$ 
is always present irrespective of no-ghost conditions 
and the rescaling of dynamical perturbations.

\bibliographystyle{mybibstyle}
\bibliography{bib}

\end{document}